\documentclass[aps,pra,superscriptaddress,amsmath,amssymb,twocolumn,showpacs]{revtex4}

\usepackage{amsmath, epsfig}
\usepackage{amssymb}
\usepackage{mathrsfs}
\usepackage{hyperref}
\usepackage[latin1]{inputenc}
\usepackage{color}

\usepackage{cancel}
\usepackage{soul}
\usepackage{ulem}
\usepackage{amsfonts}
\usepackage{amstext}
\usepackage{graphicx}
\usepackage{amscd}
\usepackage{amsmath}
\usepackage{mathrsfs}
\usepackage{enumerate}

\newcommand{\llangle}{\langle \hspace{-0.2em} \langle}
\newcommand{\rrangle}{ \rangle \hspace{-0.2em}  \rangle}
\newcommand{\transp}{\text{\textsc{t}}}
\newcommand{\tr}{\mathrm{Tr}}
\renewcommand{\eqref}[1]{Eq.~(\ref{#1})}
\newcommand{\ket}[1]{\vert #1\rangle}
\newcommand{\bra}[1]{\langle #1\vert}
\newcommand{\Ket}[1]{\vert #1 \rrangle}
\newcommand{\Bra}[1]{\llangle #1\vert}

\begin{document}

\title{Quantum synchronization as a local signature of  super- and subradiance }

\author{B. Bellomo}
\affiliation{Institut UTINAM - UMR6213,  CNRS,  Observatoire des Sciences de
l'Univers THETA,  Universit\'{e} Bourgogne Franche-Comt\'{e}, F-25000 Besan\c con,
France}

\author{G. L. Giorgi}
\affiliation{
IFISC (UIB-CSIC), Instituto de F\'{i}sica Interdisciplinar y Sistemas Complejos (Universitat de les Illes Balears-Consejo Superior de Investigaciones Cient\'{i}ficas), UIB Campus, E-07122 Palma de Mallorca, Spain}

\author{G. M. Palma}
\affiliation{NEST,   Istituto  Nanoscienze-Consiglio Nazionale delle Ricerche and  Dipartimento  di  Fisica
e  Chimica, \\
Universit\`{a} degli Studi di Palermo, Via Archirafi 36, I-90123 Palermo,
Italy}

\author{R. Zambrini}
\affiliation{
IFISC (UIB-CSIC), Instituto de F\'{i}sica Interdisciplinar y Sistemas Complejos (Universitat de les Illes Balears-Consejo Superior de Investigaciones Cient\'{i}ficas), UIB Campus, E-07122 Palma de Mallorca, Spain}


\begin{abstract}

We study the relationship between the collective phenomena of super- and subradiance and
spontaneous synchronization of  quantum systems.
To this aim we revisit the case of two detuned qubits interacting through a pure dissipative
bosonic environment, which contains the minimal ingredients for our analysis.
By using the Liouville formalism, we are able to find analytically the ultimate connection between these phenomena.
We find that dynamical synchronization is due to the presence of long standing coherence
between the ground state of the system and the subradiant state. We finally show that,
under pure dissipation, the emergence of spontaneous synchronization and of subradiant
emission occur on the same time scale. This reciprocity is broken in the presence of dephasing noise.

\end{abstract}


\maketitle

\section{Introduction}
Superradiance, the collective emission of a sample of atoms, is one of the most celebrated
examples of cooperative quantum phenomena, while synchronization is a widespread and paradigmatic emergent behavior in complex systems
(see \cite{Haroche} and \cite{Pikovsky}, respectively, for comprehensive reviews).
In the most idealized  form of superradiance, as first described by Dicke in his seminal paper \cite{Dicke},
the atoms identically interact with the electromagnetic field, a situation encountered, e.g.,
when these are confined in a region of space smaller than the wavelength of the resonant modes.
When the initial atomic state is a highly entangled symmetric state,
the emission is cooperative and takes place in  a rapid burst at a rate
proportional to the square of the number of atoms.  This is a clear signature of
the presence of constructive quantum interference. If the emission starts from the state in which all the atoms are in their
excited state, the collective atomic state acts as an antenna amplifying at a
macroscopic level the quantum field fluctuation that triggers the emission.

 Superradiance in the presence of a small
number of emitters \cite{Garraway} has been observed by employing trapped ions \cite{devoe,eschner} and superconducting qubits \cite{vanloo}.
Superradiant emission has been shown to be a useful resource towards the realization of single-photon sources
\cite{single}, quantum memories \cite{memories},  and laser cooling \cite{cooling}.
In particular, one-dimensional quantum electrodynamics systems are very promising to observe
superradiant emission, as the interaction
 can be maintained over large distances \cite{vanloo,mlynek,goban}. Furthermore, such a
 flexible setup makes it possible to explore different coupling regimes, and then to study the transition
 from the case of independent emitters to the collective-decay limit. The counterpart of superradiance, that is,
 subradiance, is significantly harder to observe experimentally, as subradiant states take place over longer time scales and require protection from any possible source of local noise. It was also experimentally observed in the case of a pair of emitters  \cite{devoe} and more recently
 in a large system of resonant scatterers in an extended and dilute cold-atom sample \cite{subradiance}.

The aim of this paper is to discuss the connection of superradiance with spontaneous synchronization, a widespread phenomenon of adjustment
of the dynamical rhythms of \textit{different} interacting systems \cite{Pikovsky}. This emergent phenomenon has been largely studied in classical regimes and multidisciplinary contexts.
It has been predicted in the quantum regime, in several platforms including harmonic networks \cite{syncHO},
mechanical resonators \cite{mec},  and atomic and ionic setups \cite{julia,zhu,hush,giorgi,syncprobe}.
Different approaches have been considered to frame quantum synchronization  referring to both local and global, e.g., quantum correlation,  measures
(see \cite{chapter} and references therein).
For the purpose of this work, we address the collective quantum phenomenon of super. and subradiance
 to establish a connection with spontaneous synchronization between the local dynamics of the atoms;
we therefore consider spontaneous synchronization as the adjustment to a common rhythm of local
observables as in Refs. \cite{giorgi,syncprobe}.

We mention that both synchronization and superradiance have also
been considered in the presence of external driving.
Driven synchronization, or entrainment, in the quantum regime has been considered, for instance, in
Ref. \cite{entrain}. On the other hand,
superradiance in the presence of driving has been identified as lasing with an extremely narrow linewidth
\cite{narrow}, with a collective emission with an extremely
high degree of phase coherence \cite{bohnet} and quantum phase locking \cite{hollande2014}. It can 
be observed in general in incoherently pumped arrays of  strongly interacting atomic
quantum dipoles \cite{zhu}. Indeed, the emission takes place in rapid bursts
with large fluctuations in the emission delay time among the various
experimental observations \cite{Palma1}. Further features in the superradiant
emissions that are a signature of synchronization take place when the emission
is due to a sample of two different, detuned atomic species
\cite{Palma2,Vaglica}.

The aim of this work is to consider both superradiance and spontaneous synchronization
taking place during dynamical relaxation into a common environment, in the absence of external driving
and in the presence of detuning between the atoms.
In order to get a full analytic treatment,  we consider within the Liouville formalism the case of two detuned
two-level systems in the presence of dissipation induced by a thermal
environment, displaying both superradiant and subradiant emission and
dynamical synchronization.
We note that, although collective effects are more prominent in the presence of a large number
of emitters,  most
features are already present in the  decay of just two atoms coupled via dipole interactions \cite{Tanas1,Palma3,Tanas2}.
We will show that, in
the presence of pure dissipation, the two phenomena are different manifestations
of a sole property of the model. Indeed, dynamical synchronization is due to
the presence of coherence between the ground state and the subradiant state.
Thus, this kind of synchronization has a pure quantum origin, as it directly
relies on  quantum interference.


The paper is organized as follows. In Sec. \ref{model}, we introduce the model
and the master equation used to solve the dynamics. In Sec. \ref{dmev} the
evolution of the system is studied by employing the so-called Liouville
representation for the density matrix. In Sec. \ref{Sec: synchronization} we
introduce quantum synchronization and quantify it in our model. Super- and
subradiance are studied in Sec.  \ref{subr} and compared to synchronization. The
case of purely dephasing noise is the subject of Sec. \ref{depnoise}. Finally,
a summary given in Sec. \ref{conclusions}.

\section{Model}\label{model}

Our system $S$ consists of two  atoms that interact with a common bosonic
environment $E$. We will show that this model, already considered in several works \cite{Agarwal1974,ficek},
allows for a fully analytical description of the origin of the connection between  the phenomena of  synchronization and
of super- and subradiance.
The two atoms (described by qubits $1$ and $2$) have different
frequencies $\omega_1$ and $\omega_2$ and their free Hamiltonian  is
($\hbar=1$)
\begin{equation}
H_S =\frac{\omega_1}{2} \sigma_1^z +\frac{\omega_2}{2}
\sigma_2^z,
\end{equation}
where $\sigma^z_{1,2}=\ket{e}\bra{e}-\ket{g}\bra{g}$ and $|g\rangle$ and $|e \rangle$
indicate, respectively, the ground and excited states of each atom.
The eigenstates of the free
Hamiltonian $H_S$ form the so-called decoupled basis $\{\ket{ee}, \ket{eg},
\ket{ge}, \ket{gg}\}$ with corresponding energies $\{\omega_0, \delta, -\delta,
-\omega_0\}$, where $\omega_0=(\omega_1+\omega_2)/2$ is the average
frequency
and $\delta=\omega_1-\omega_2$   is the detuning between atoms.

The total Hamiltonian of the atoms interacting with the environment is
$H_T=H_\mathrm{S}+H_\mathrm{E}+H_\mathrm{I}$, where $H_\mathrm{E}= \sum_k
\omega_k a_k^\dag a_k$ corresponds to the radiation environment and
$H_\mathrm{I}=- \sum_i \mathbf{\hat{d}}_i \cdot \mathbf{E}_i$  is the
interaction Hamiltonian between the two atoms ($\mathbf{\hat{d}}_i
=\mathbf{d}_i\sigma^x_i$, with $\mathbf{d}_{i}=\bra{e}\mathbf{\hat{d}}_i
\ket{g}$, is the dipole operator of atom $i$th) and the environmental field
$\mathbf{E}_i=\sum_k \mathbf{\mathcal{E}}_k^i (a_k+a_k^\dag)$.
The interaction Hamiltonian can be thus put in the form \cite{Agarwal1974}
\begin{equation}
H_I =  \sum_i  \sigma^x_i   \sum_k  g_k^i (a_k+a_k^\dag)    ,
\end{equation}
where $g_k^i=- \mathbf{d}_i  \cdot  \mathbf{\mathcal{E}}_k^i  $.

\subsection{Master equation}

In the limit of weak system-bath coupling and for zero temperature, the
dynamics of the reduced density matrix of the two atoms $\rho$ can be studied by
performing the   Born-Markov and the secular approximations, noting  that, in the limit of small detuning $\delta \ll\omega_0$, also
terms oscillating with
frequency $\omega_1-\omega_2$ must be kept. The master equation reads
\cite{Agarwal1974, BreuerBook}
\begin{equation} \label{MasterEquation}
\dot \rho = -i\left[H_S+H_{LS}, \rho \right] + \sum_{ij}  \gamma_{ij} \left(
\sigma_i^- \rho  \sigma_j^+ -\frac{1}{2} \left\{ \sigma_j^+ \sigma_i^-,\rho
\right\} \right),
\end{equation}
where $\sigma_i^{+}=\ket{e}\bra{g}$ and $\sigma_i^{-}=\ket{g}\bra{e}$.  In the
following we use the notation
$\gamma_{ii}\equiv\gamma_{i}$ and define $\gamma_0=(\gamma_1+\gamma_2)/2$
while
$\gamma_{12}=\gamma_{21}$ is assumed to be real (its maximal absolute  value is given by
$\sqrt{\gamma_1 \gamma_2}$).
The Lamb-shift Hamiltonian is
\begin{equation}\label{LambShift}
 H_{LS}=s_1 \sigma^z_1  +s_2  \sigma^z_2 +
s_{12}\left(\sigma_1^- \sigma_2^+ + \sigma_1^+ \sigma_2^-\right).
\end{equation}
The local  shifts  $s_i$
lead to irrelevant  renormalization of  the natural frequencies $\omega_1$
and $\omega_2$, and $s_{12}$ is assumed to be real.

The different parameters appearing in the master equation are directly
connected
to the correlations functions of the environment, and thus to its spectral
density.
In particular, the damping coefficients $\gamma_{ij}$
depend on the interaction Hamiltonian parameters and on the atomic transition frequencies.
Their explicit form in the vacuum electromagnetic field can be found, for instance, in Eqs. (31-33) of Ref. \cite{Tanas2}.
Different physical configurations  correspond to different  regimes in the parameter space. As an example, a crucial factor to determine whether or not the super- and subradiant regime is achieved or not is the distance between the atoms.
For large separations, the coupling terms  $\gamma_{ij}$, with $i\neq j$, are negligible and start being relevant when the separation is  of the same order of magnitude as the resonant wavelength. These parameters attain their maximal
values in the small-sample limit. 
Henceforth, in order to explore different regimes, we assume to be
able
to tune the values of these parameters freely.

\section{Density matrix evolution in the Liouville representation}\label{dmev}

The dynamics of the system can be studied by adopting the  so-called
Liouville representation of the density matrix. In this representation, the
master equation (\ref{MasterEquation}) is mapped into a set of linearly
coupled differential equations for the elements of the reduced density
matrix of the two atoms: $|\dot \rho_t \rrangle = \mathcal{L} |\rho_t
\rrangle$,
where  $|\rho_t \rrangle$ is a vector in the  Hilbert-Schmidt space, $\mathcal
H
= \mathbb C^{16}$:
\begin{equation}\label{DefinitionRL}
\rho = \sum_{i,j=1}^4 \rho_{ij} |i\rangle \langle j| \xrightarrow{HS} \Ket{\rho}
= \sum_{i,j=1}^4 \rho_{ij}  \Ket{ij } ,
\end{equation}
where $ \Ket{ij }= |i\rangle \otimes |j \rangle$. The inner product in the
Hilbert-Schmidt space is defined by $\llangle \tau  \Ket{ \rho } =
\tr(\tau^\dagger \rho)$.

As shown in the Appendix, the total
Liouvillian $\mathcal{L}$ turns out to be the direct sum of five terms
$\mathcal{L}=\bigoplus_{\mu} \mathcal{L}_\mu$, $\mu \in \{a, b, c, d, e\}$.
Thus, $\mathcal H$ can be decomposed into 5 independent subspaces, $\mathcal H
=
\bigoplus_{\mu} \mathcal{H}_\mu$, whose dimensions are $6$, $4$, $4$, $1$ and $1$.  The
form of each $\mathcal{L}_\mu$  and the corresponding right ($\Ket{
\tau^\mu_i}$) and left ($\Bra{ \bar{\tau}^\mu_i}$) eigenvectors, together with
the eigenvalues, is explicitly
reported in the Appendix. We stress that these eigenvectors are
not constrained to represent valid states.  In fact, as we will see
later, the only physical state that is also an
eigenvector of  $\mathcal{L}$ is the ground state, while the totally excited
state is the only physical eigenvector of $\mathcal{L}^\dag$. This implies
that,
for instance, a generic  $\Ket{ \tau^\mu_i}$ cannot be taken as the initial state.
On the other hand, the dynamics of any density matrix can always be described
as
a superposition of such nonphysical objects whose  linear combination is
constrained to be a density matrix.

Within this representation, a key role is played
by $\mathcal{L}_a$ being $\mathcal{H}_a$ generated by the projectors of the
four
states forming the decoupled basis and by $\ket{eg}\bra{ ge}$ and
$\ket{ge}\bra{eg}$.
As shown in the Appendix, Sec. \ref{Eigenvectors and eigenvalues}, the right (R)
eigenvectors of
$\mathcal{L}_a$ [\eqref{EigensystemL0}] are simply expressed  in terms of two
nonorthogonal states, $\ket{S_R}$ and $\ket{A_R}$, given  by
\begin{eqnarray}\label{Stilde e Atilde}
 \ket{S_R} &=& \frac{\alpha_S \ket{eg} +\ket{ge}}{\sqrt{1+|\alpha_S|^2}} , \quad
  \alpha_S =\frac{\frac{\gamma_1-\gamma_2}{2}+i \delta +V}{\gamma_{12}+ i \,
2\,
s_{12}},
\nonumber \\
  \ket{A_R} &=& \frac{\alpha_A \ket{eg} +\ket{ge}}{\sqrt{1+|\alpha_A|^2}} ,
\quad
  \alpha_A =\frac{\frac{\gamma_1-\gamma_2}{2}+i \delta -V}{\gamma_{12}+i \, 2\,
s_{12}}, \nonumber \\
 \mathrm{being} & &\!\!\!\!\!\! V = \sqrt{(\gamma_{12}+i \,2 \,  s_{12})^2 +
\left(\frac{\gamma_1-\gamma_2}{2}+i \delta \right)^2}.
\end{eqnarray}
This parameter $V$ characterizes all the decay rates of the system
(we use the notation  $V_\mathrm{r}=\mathrm{Re} [V]$ and
$V_\mathrm{i}=\mathrm{Im} [V]$) appearing in the eigenvalues of
$\mathcal{L}_a$
[\eqref{EigensystemL0}]. In the symmetric configuration, when
$\gamma_1=\gamma_2$ and $\delta=0$, $V$ simply reduces to  $\gamma_{12}+i 2
s_{12}$. On the other hand, the left (L) eigenvectors of $\mathcal{L}_a$
(\eqref{LeftL0}) are
given in terms of the states
\begin{eqnarray}
\ket{S_L}=\ket{S_R}^*, ~~  \ket{A_L}=\ket{A_R}^*.
\end{eqnarray}
The property $\alpha_S \, \alpha_A=-1$  implies that
$\langle S_L \ket{A_R}=\langle A_L \ket{S_R}=0$.

Notice that, changing $\gamma_{12}$ into $-\gamma_{12}$ (this quantity can also assume negative values \cite{lehmberg}) the same spectrum of the Liouvillian is obtained provided that the values of $\gamma_{1}$ and $\gamma_{2}$ are also interchanged. Furthermore, as a reflection of the Liouvillian symmetry, in the case $\gamma_{1}=\gamma_{2}$ and $\delta=0$, the subradiant state switches to the superradiant one and vice versa.
On the other hand, if $\gamma_{1}=\gamma_{2}$ but $\delta\neq 0$, $\gamma_{12}\to-\gamma_{12}$ implies $\ket{S_R}\to \ket{S_L}^*$ and $\ket{S_L}\to \ket{S_R}^*$.

\begin{figure}[t]
  \centering
  \includegraphics[width=0.47\textwidth]{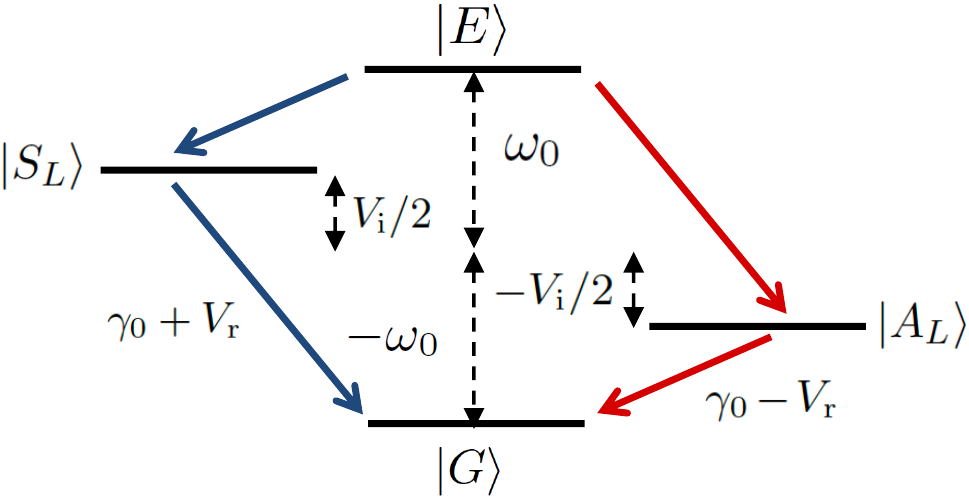}
  \caption{Schematic representation of  two independent channels connecting the
excited state $\ket{E}=:\ket{ee}$ (energy equal to $\omega_0$) to the
ground state
$\ket{G}=:\ket{gg}$ (energy equal to $-\omega_0$) [see \eqref{rate
equation}]. The channel
passing through the state $\ket{S_L}=\ket{S_R}^*$ [whose oscillations are
associated with  $V_\mathrm{i}/2$; see \eqref{EigensystemLX1}]
is characterized by a decay rate towards the ground state equal to $\gamma_0 +
V_\mathrm{r}$  ($\gamma_0=\frac{\gamma_1+\gamma_2}{2}$). The channel
passing
through the state $\ket{A_L}=\ket{A_R}^*$ (whose oscillations are associated with
$-V_\mathrm{i}/2$)  is characterized by a decay rate towards the ground
state
equal to $\gamma_0 - V_\mathrm{r}$. For simplicity, the  channels
connecting
$\ket{E}$  and $\ket{G}$ through the coherences between  $\ket{S_L}$ and
$\ket{A_L}$ are not shown.}\label{FigTrans}
\end{figure}

The states $\ket{S_R}$ and $\ket{S_L}$, and $\ket{A_R}$ and $\ket{A_L}$,
appear to be a natural extension of the symmetric $\ket{S} = (\ket{ge}+
\ket{eg} )/\sqrt{2}$, and antisymmetric  $\ket{A} = (\ket{ge}- \ket{eg}
)/\sqrt{2}$, states, and they reduce to them in the symmetric configuration
($\gamma_1=\gamma_2$ and $\delta=0$). Their relevance  in the two-qubit
dynamics  is twofold. On the one hand,  $\ket{S_R}$ and $\ket{A_R}$ enter 
the expression of the right eigenvectors  $\Ket{ \tau^a_i}$, so that they can
naturally  be used to get simple expressions for the evolution of the 
density-matrix operator [see \eqref{rhot]}. On the other hand, $\ket{S_L}$ and
$\ket{A_L}$ enter  the expression of the left eigenvectors  $\Ket{
\bar{\tau}^a_i}$ [see  \eqref{eig}]
so the rate equations for the evolution of the populations
of the density matrix have the simple form reported in \eqref{rate
equation}.
This is schematically represented in Fig. \ref{FigTrans}, which shows that the
states $\ket{S_L}$ and $\ket{A_L}$ connect the excited state to the ground
state
through two independent channels. In particular, the decay rates from
$\ket{S_L}$ and $\ket{A_L}$ to the ground state are given, respectively,  by
$\gamma_0 + V_\mathrm{r}$ and $\gamma_0 - V_\mathrm{r}$, natural
extensions of
the decay rates in the symmetric configuration, $\gamma_0 + \gamma_{12}$ and
$\gamma_0 - \gamma_{12}$. The
states $\ket{S_L}$ and $\ket{A_L}$  indeed play a role analogous to the one
played
by the superradiant state $\ket{S}$ and by  the subradiant state $\ket{A}$ in
the symmetric case.

We now show that the evolution of the density-matrix operator is simply
expressed in terms of the right eigenvectors of $\mathcal{L}$. By using the decomposition for the identity operator in the  Hilbert-Schmidt space
reported in \eqref{identity} and $|\dot \rho_t \rrangle = \mathcal{L} |\rho_t
\rrangle$, one can show that, given an arbitrary initial state $\Ket{\rho_0}$,
the evolved density matrix can be written (apart from the case of degeneracy in
the  spectrum of the Liouvillian, as in the case of the Dicke limit when
$\gamma_1=\gamma_2=\gamma_{12}$ and $\delta=0$) as
\begin{equation}\label{rhot}
  \Ket{\rho_t}=\sum_{\mu}\sum_{i} p_{0\, i}^{\mu}\, \Ket{ \tau^\mu_i} \,
\mathrm{e}^{\lambda_i^\mu t}, \:\: \mathrm{being} \:\:  p_{0\,
i}^{\mu}=\frac{\llangle \bar{\tau}^{\mu}_i \Ket{ \rho_0}}{\llangle
\bar{\tau}^{\mu}_i\Ket{\tau^{\mu}_i}},
\end{equation}
where $\mu$ runs over the five subspaces and $i$ between 1 and  ${\rm
dim}(\mathcal{H}_\mu)$. Equation (\ref{rhot}) is the main tool we are going to
use in
the following sections to make a connection between the phenomena of synchronization
and of super- and subradiance.

Before doing that, we consider a couple of examples that
will be useful in the rest of this paper. Under particular
initial conditions the dynamics is governed by only one decay rate
(different from zero). If we start from $\Ket{\rho_0}=\Ket{A_R A_R}$ (i.e., we
prepare the system in the pure state $\ket{A_R}$), one can see from
\eqref{EigensystemL0} that only two right eigenvectors of $\mathcal{L}_A$
contribute to its dynamics, $\Ket{\rho_0}=\Ket{\tau^a_1}+\Ket{\tau^a_6}$. It
follows that, in \eqref{rhot},  $p_{0\, 1}^{a}=p_{0\, 6}^{a}=1$ (the others
$p_{0\, i}^{\mu}$ are equal to zero), so   at time $t$ we have
\begin{eqnarray}\label{rhoAt}
  \Ket{\rho_t}&=& \Ket{\tau^a_1}+\Ket{\tau^a_6} \mathrm{e}^{\lambda_6^a t}  \\
&=&   \Ket{GG} \left[1-\mathrm{e}^{-\left(\gamma_0- V_\mathrm{r}\right)t
}\right]+
  \Ket{A_R A_R} \mathrm{e}^{-\left(\gamma_0- V_\mathrm{r}\right)t}\,
\nonumber
.
\end{eqnarray}
Equation (\ref{rhoAt}) clearly shows that in this case the vector $\Ket{A_R A_R}$
is
dynamically coupled only to $\Ket{GG}$, its decay rate being given by
$\gamma_0-
V_\mathrm{r}$.
Its role is in this sense analogous to the one played by the subradiant
antisymmetric state  when $\gamma_1=\gamma_2$ and $\delta=0$.

Analogous considerations hold for the  vector  $\Ket{S_R S_R}$. If we prepare
the system in $\Ket{\rho_0}=\Ket{S_RS_R}$, this implies, using
\eqref{EigensystemL0}, that $\Ket{\rho_0}=\Ket{\tau^a_1}+\Ket{\tau^a_5}$. It
follows that, $p_{0\, 1}^{a}=p_{0\, 5}^{a}=1$, so  the evolution takes the
form
\begin{eqnarray}\label{rhoSt}
  \Ket{\rho_t}&=& \Ket{\tau^a_1}+\Ket{\tau^a_5} \mathrm{e}^{\lambda_5^a t} \\
&=
&  \Ket{G G} \left[1-\mathrm{e}^{-\left(\gamma_0+
V_\mathrm{r}\right)t}\right]+
  \Ket{S_R S_R} \mathrm{e}^{-\left(\gamma_0+ V_\mathrm{r}\right)t}\, .
\nonumber
\end{eqnarray}
The role of the vector $\Ket{S_RS_R}$ is thus analogous to the one played by
the
superradiant symmetric state in the case when $\gamma_1=\gamma_2$ and
$\delta=0$.

\section{Synchronization}\label{Sec: synchronization}

This section is devoted to the quantum synchronization phenomenon in our
two-qubit system. As we deal with a system relaxing towards a thermal state (at zero temperature, i.e., the ground state),
the emergence of synchronization
is assessed in a transient (preasymptotic) regime
\cite{syncHO,giorgi}. We focus on the case of spontaneous synchronization, emerging during the dynamics in the absence of pumping sources at the atoms' locations.
Quantum spontaneous synchronization can be quantified through different
measures,
and local and global indicators and  correlations
have been used for this purpose, as reviewed in Ref. \cite{chapter}. 
A global indicator for quantum synchronization in an atomic context has been considered, for instance, in Ref. \cite{hollande2014}, and
identified as the absence of a relative oscillation in the correlation at different times of two driven-atom clouds. Interestingly,
under proper conditions, this correlation can be accessed experimentally through optical measurements. 

Here we are interested in comparing the (collective) phenomenon of superradiance
with the synchronous evolutions of the atoms' (local) dynamics of either populations or coherences.
To this aim, we compute the evolution of the average of arbitrary single-atom
operators $O_k$ ($k=1,2$), looking at their respective oscillatory (generally
irregular) dynamics.  We
consider the expectation values $\langle O_1\rangle_t$ and $\langle
O_2\rangle_t$,  where $\langle \bullet \rangle_t= \mathrm{Tr}(\bullet \rho_t)$
indicates the average of an arbitrary operator $\bullet$ at time $t$.
Synchronization between these local observables is quantified through
the Pearson
correlation coefficient, a  widely used  measure of the degree
of linear dependence between two variables. Given two time-dependent variables
$A_1$ and $A_2$, the Pearson factor $C$ is
\begin{equation}\label{eq:pears}
 C_{A_1(t),A_2(t)}(\Delta t)=\frac{\int_{t}^{t+\Delta
t}[A_1(t)-\bar{A_1}][A_2(t)-\bar{A_2}]dt^\prime}
 {\sqrt{ \prod_{i=1}^2\int_{t}^{t+\Delta t}[A_i(t)-\bar{A_i}]^2 dt^\prime  }},
\end{equation}
where $ \bar{A_i}=\frac{1}{\Delta t}\int_{t}^{t+\Delta t}A_idt^\prime $
and $A_i$ are expectation values of quantum operators. This synchronization
measure is evaluated in time,
considering a
sliding window of length $ \Delta t$.
 As a consequence of the definition,
$C$ gives a value between $+1$ and $-1$,  where $+1$ ($-1$ ) indicates the
presence of in-phase (antiphase) synchronization and $0$  the absence of any
synchronization. This measure can be
generalized to the case of certain time delay between the local dynamics
allowing for a phase shift,
i.e. embodying a time delay $\delta t$ in one of the two signals within the
definition of
$C$ as discussed in Ref. \cite{chapter}. In this way,
$C_{A_1(t),A_2(t+\delta t)}(\Delta t)=+1$ would indicate perfect
time-delayed synchronization with delay $\delta t$.

The dynamical study conducted using the Liouville representation allows us to
establish how the time scale of the phenomenon of synchronization depends
analytically on the physical parameters appearing in the master
equation.
Any pair of arbitrary local operators $O_k$, $k=1,2$,  can be
written in the single-atom basis (Bloch representation)
$\{\sigma^x_k, \sigma^y_k, \sigma^z_k, I_k^d\}$:
\begin{equation}\label{OperatoreArbitrario}
  O_k= a^x_k \sigma^x_k + a^y_k \sigma^y_k + a^z_k \sigma^z_k + a^d_k I_k^d.
\end{equation}

By looking at the decomposition of the Liouvillian operator (see the Appendix, Sec.
\ref{Ldecomp}),  we find that only the right eigenvectors of $\mathcal{L}^b$
and $\mathcal{L}^c $  contribute to the average of $\sigma^x_k $ and $\sigma^y_k
$, while only the ones  of $\mathcal{L}^a$ contribute to the average of
$\sigma^z_k $. It follows that, using \eqref{rhot}, we have
\begin{eqnarray}\label{MediaA}
   \langle O_k\rangle_t &=& a^x_k  \langle \sigma^x_k \rangle_t + a^y_k \langle
\sigma^y_k \rangle_t +  a^z_k  \langle  \sigma^z_k \rangle_t + a_k^d , \qquad
\mathrm{where}\nonumber \\
   \langle \sigma^x_k \rangle_t
    &=& \sum_{i} 2 \left| p_{0\, i}^{b}\,   \langle \tau^b_i  \rangle_{xk}
\right| \, \mathrm{e}^{\mathrm{Re}(\lambda_i^b) t} \,
   \cos \left[\mathrm{Im}(\lambda_i^b) t+ \varphi_{i, xk}^b \right],\nonumber
\\
   \langle \sigma^y_k \rangle_t &=& \sum_{i} 2 \left| p_{0\, i}^{b}\,   \langle
\tau^b_i  \rangle_{yk}  \right| \, \mathrm{e}^{\mathrm{Re}(\lambda_i^b) t} \,
   \cos \left[\mathrm{Im}(\lambda_i^b) t+ \varphi_{i, yk}^b \right],\nonumber
\\
   \langle \sigma^z_k \rangle_t &=& \sum_{i} p_{0\, i}^{a}\,   \langle \tau^a_i
 \rangle_{zk}   \, \mathrm{e}^{\lambda_i^a t},
\end{eqnarray}
with $\langle \tau^\mu_i  \rangle_{\nu k}= \mathrm{Tr}(\sigma^\nu_k
\tau^\mu_i)$ and $\varphi_{i, \nu k}^b = \mathrm{Arg} \left(p_{0\, i}^{b}\,
\langle \tau^b_i  \rangle_{\nu k} \right)$, for $\nu=x ,y, z$, and where we
used $\lambda_i^c=\lambda_i^{b\,*}$, $\langle \tau^c_i  \rangle_{\nu k}=\langle
\tau^b_i  \rangle_{\nu k}^*$ and $p_{0\, i}^{c}=p_{0\, i}^{b\,*}$.
Here $ \tau^\mu_i$ is the matrix obtained by mapping the eigenstate
$\Ket{\tau^\mu_i} $ back onto the Hilbert space ($\tau^\mu_i \xrightarrow{HS}
\Ket{\tau^\mu_i}$). We recall that $\tau^\mu_i $ is not necessarily a density
matrix, while any $\langle \sigma^\nu_k \rangle_t$ is well defined at any time.

Equation (\ref{EigensystemL0}) implies that only one frequency, equal to
$V_{\mathrm{i}}$, enters  $\langle \sigma^z_k \rangle_t$. The term
oscillating with this frequency decays with a rate $\gamma_0$.
As for the averages of $\sigma^x_k$ and $\sigma^y_k$, each of them contains
four
oscillating terms. In the worst-case scenario, when all  four terms
contribute to these averages, the ratios between the instantaneous amplitudes
of
these oscillating terms are proportional to
$\mathrm{e}^{\left[\mathrm{Re}(\lambda_i^b) -
\mathrm{Re}(\lambda_j^b)\right]t}$. By using \eqref{EigensystemLX1} one sees
that the terms with smallest decay rates are the one with $i=4$ and $i=3$. The
long-lasting ratio between the various terms and the one with smallest decay
$(i=4)$ decays as $\mathrm{e}^{\left[\mathrm{Re}(\lambda_3^b) -
\mathrm{Re}(\lambda_4^b)\right]t}=\mathrm{e}^{-V_\mathrm{r}t} $. In a time
scale
given by several units of $V_\mathrm{r}^{-1}$, we expect that only one term
still
contributes to the averages of $\langle \sigma^x_k \rangle_t$ and $\langle
\sigma^y_k \rangle_t$. This is the time-scale separation enabling
the emergence of spontaneous synchronization
\cite{syncHO,giorgi}. On the other hand, the decay rate of the oscillating
term
in $\langle \sigma^z_k \rangle_t$ is always larger than $V_{\mathrm{r}}$.
The time scale for the occurrence of synchronization is thus characterized by
a
parameter $\kappa_S$ equal to the real part of $V$, while the frequency of the
long lasting term, giving the frequency of the synchronized oscillations,  is
related to  the imaginary part of $V$:
\begin{equation}\label{Sincronizzazione}
    \kappa_S=V_{\mathrm{r}}, \qquad
    \nu_S=\omega_0 - \frac{1}{2}V_{\mathrm{i}}  .
    \end{equation}
The real and imaginary parts of $V$ obey $V_{\mathrm{r}}V_{\mathrm{i}}=2
\gamma_{12} s_{12} + \delta (\gamma_1-\gamma_2)/2$.

\begin{figure}[t!]
  \centering
  \includegraphics[width=0.45 \textwidth]{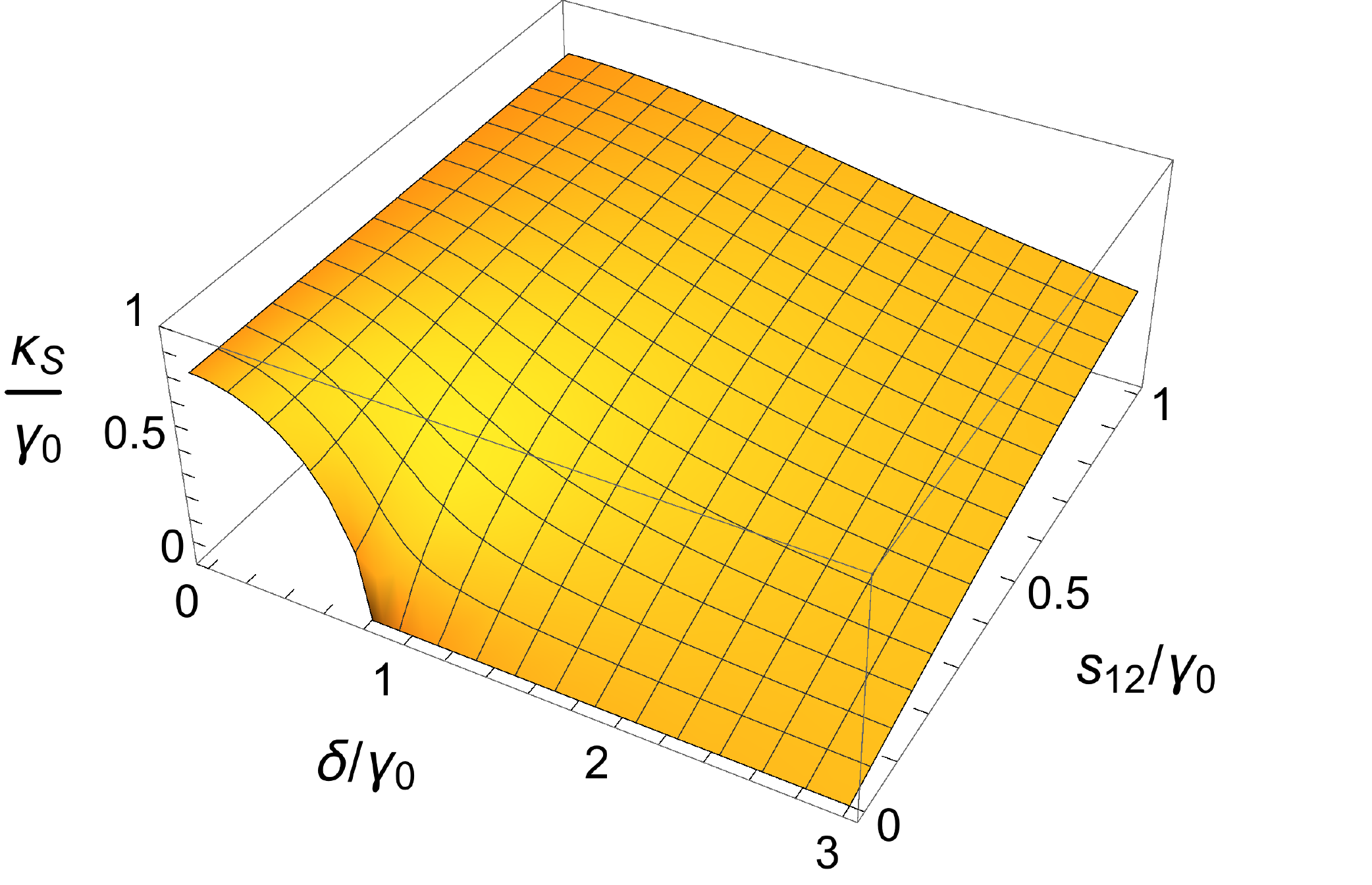}\\
  \caption{Plot of $\kappa_S$ as a function of $\delta$ and $s_{12}$ in units of
$\gamma_0$. The values of the other  parameters are $\gamma_1 =\gamma_2= \gamma_0$ and
$\gamma_{12} =0.8  \gamma_0$. }\label{FigSP}
\end{figure}

The spontaneous emergence of synchronization in a time window is then due to the long lifetime of the eigenmode $ \Ket{\tau^b_4} = \Ket{A_R G}$. In other words, the synchronization between
averages of arbitrary local operators is due
to the presence of long-standing coherences between the
ground state and the state $\ket{A_R}$, whose role in the dynamics has been already identified as analogous to the subradiant antisymmetric state. This finding clearly points out in an analytical way the connection between synchronization and subradiant emission, which will be further detailed in the next section.
 As a side note, we remark that the considerations about the exchange $\gamma_{12}\to-\gamma_{12}$ made in Sec. \ref{dmev}  imply that, under such an exchange, the frequency of synchronization would  experience a finite jump.

The above considerations allow us to determine the analytical dependence of
synchronization  on the physical parameters appearing in the master equation.
Figure \ref{FigSP} shows the inverse time of synchronization emergence
$\kappa_S$ depending on $\delta$ and on
$s_{12}$, in the case $\gamma_1=\gamma_2=\gamma_0$. In particular,
when $s_{12}=0$, $V$ reduces to $\sqrt{\gamma_{12}^2-\delta^2}$, so  for
$\delta \ge \gamma_{12}$,
$\kappa_S$ vanishes and no time-scale separation (synchronization) is found \cite{footnote}.
Decreasing the atoms' detuning, a sharp transition to synchronization
is predicted for $\delta < \gamma_{12}$. This transition to a synchronous dynamics
when decreasing the detuning is  smooth for
 $s_{12}\neq 0$, as shown in Fig. \ref{FigSP}.

\begin{figure}
  \centering
  \includegraphics[width=0.45 \textwidth]{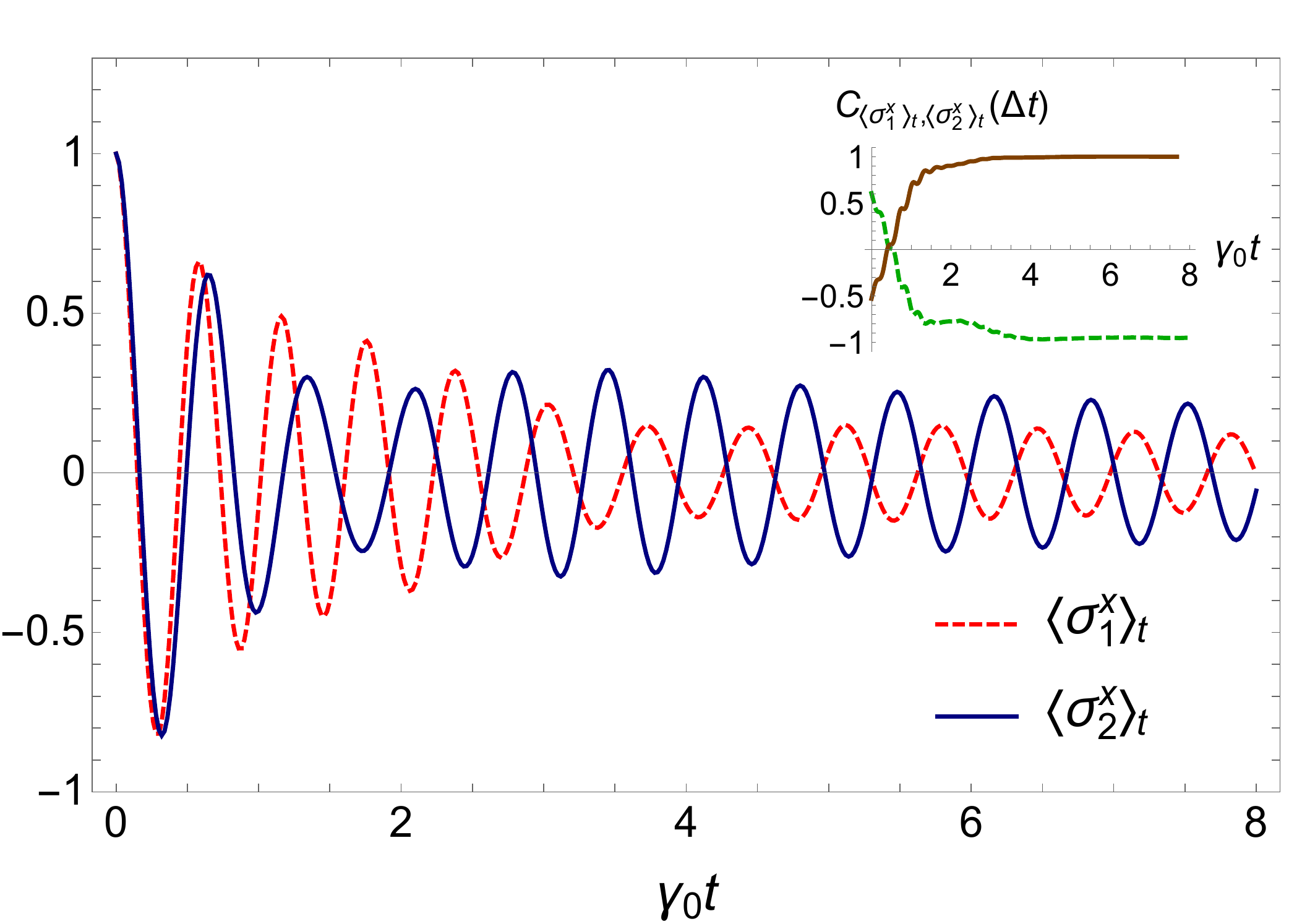}\\
  \caption{Time evolution of $\langle \sigma^x_1 \rangle_t$ (dashed line) and
$\langle \sigma^x_2 \rangle_t$ (solid line)  for the initial state
$\frac{1}{2}(\ket{e}+\ket{g})(\ket{e}+\ket{g})$.  The values of the parameters are
$\gamma_1 =1.1
\gamma_0$, $\gamma_2 =0.9 \gamma_0$, $\gamma_{12} =0.95  \gamma_0$, $s_{12}
=0.6
\gamma_0$, $\delta= \gamma_0$, and $\omega_0=10 \gamma_0$.
In this case $\kappa_S^{-1}  \gamma_0 \approx 1.2$, while the
frequency of synchronization is $\nu_S=9.26 \gamma_0$.  In the inset, we plot
the evolution of the synchronization measure $ C_{\langle
\sigma^x_1 \rangle_{t},\langle \sigma^x_2 \rangle_{t}}  (\Delta t) $ for $\Delta
t=2/\gamma_0)$   (dashed curve) and  time delayed one
  (solid line).}\label{FigSync1}
\end{figure}

In Fig. \ref{FigSync1} the emergence of spontaneous synchronization is shown in
the case when the initial state is
$\frac{1}{2}(\ket{e}+\ket{g})(\ket{e}+\ket{g})$. The plot shows that, during
the
first stage of the evolution, the dynamics of the averages $\langle \sigma^x_1
\rangle_t$  and $\langle \sigma^x_2 \rangle_t$ are characterized by several
frequencies while,  after  $t\sim\kappa_S^{-1}$, only one term
oscillating with the frequency $\nu_S$  survives.
In the inset, the time evolution of the synchronization factor $C_{\langle
\sigma^x_1 \rangle_{t},\langle \sigma^x_2 \rangle_{t}}  (\Delta t) $ is plotted as a
function
of time. The dashed curve  shows that the two averages $\langle \sigma^x_1
\rangle_t$  and $\langle \sigma^x_2 \rangle_t$  become almost antisynchronized
after some units of $\kappa_S^{-1}$. This means that while the two signals have
become almost monochromatic, their dephasing is close (but not equal) to
$180^\circ$. The
time delayed synchronization $C_{\langle \sigma^x_1 \rangle_t,\langle
\sigma^x_2
\rangle_{t+\delta t}} (\Delta t)$ is represented by the solid line. The delay $\delta t$
is the one maximizing the value of $C$.

We remark that it  may appear difficult to resolve the fast oscillation and verify the synchronization for values of $\omega_0 \gg \gamma_0$. This is the typical situation encountered in ultracold atoms. 
In this scenario the synchronization could be observed via homodyne and interferometric detection, which allows for phase detection, getting rid of the fast oscillation at the atomic transition frequency. 
On the other hand, 
there  also exist different setups where superradiance has been experimentally observed in the presence of relatively strong  decay rates. For instance, in Ref. \cite{mlynek} two transmon qubits coupled to a single coplanar waveguide resonator were considered in the strong cavity-atom coupling regime,  where radiative terms are much larger than any other competing noise effect.

\begin{figure}[t]
  \centering
  \includegraphics[width=0.35 \textwidth]{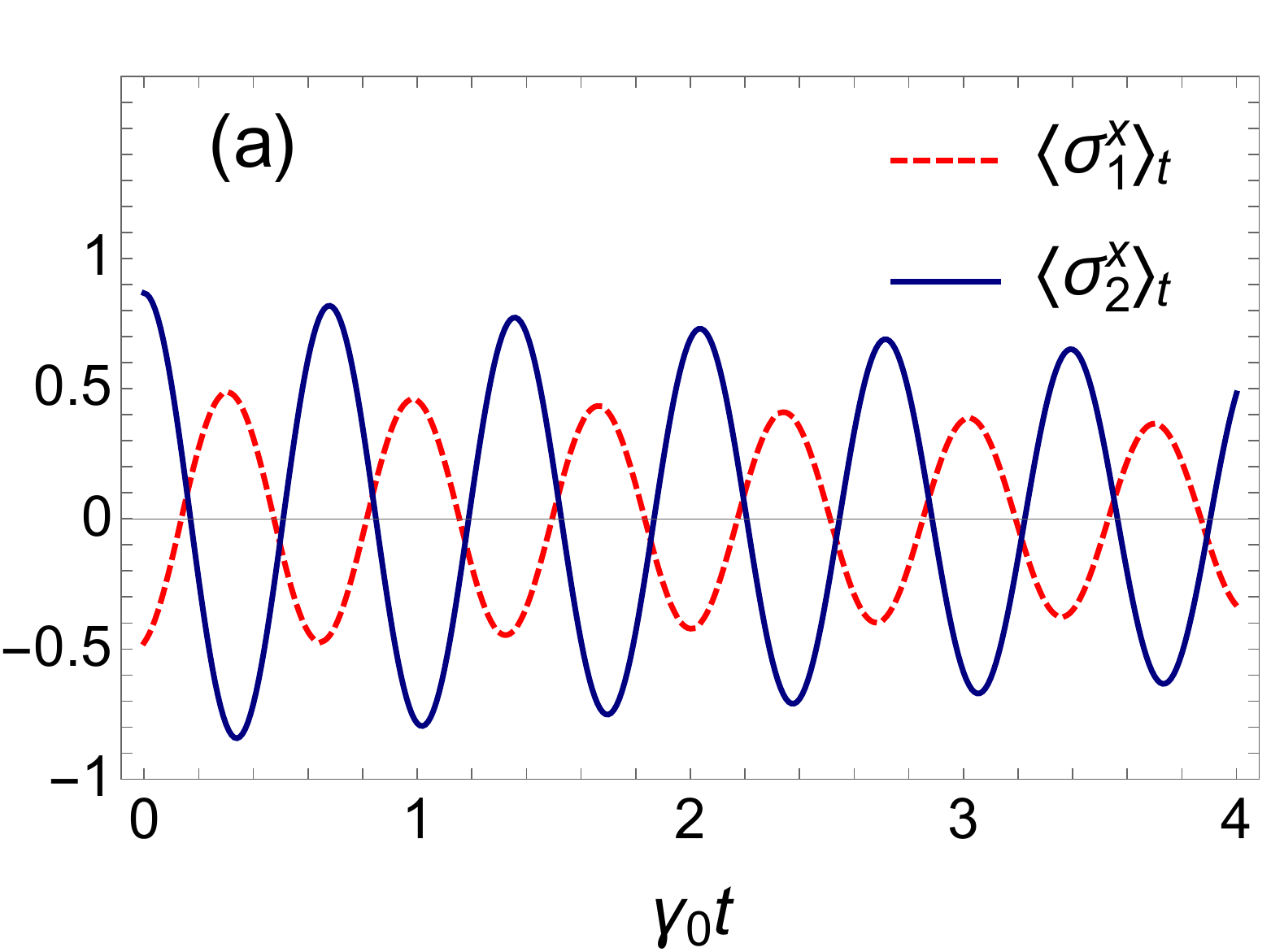}
  \includegraphics[width=0.35
  \textwidth]{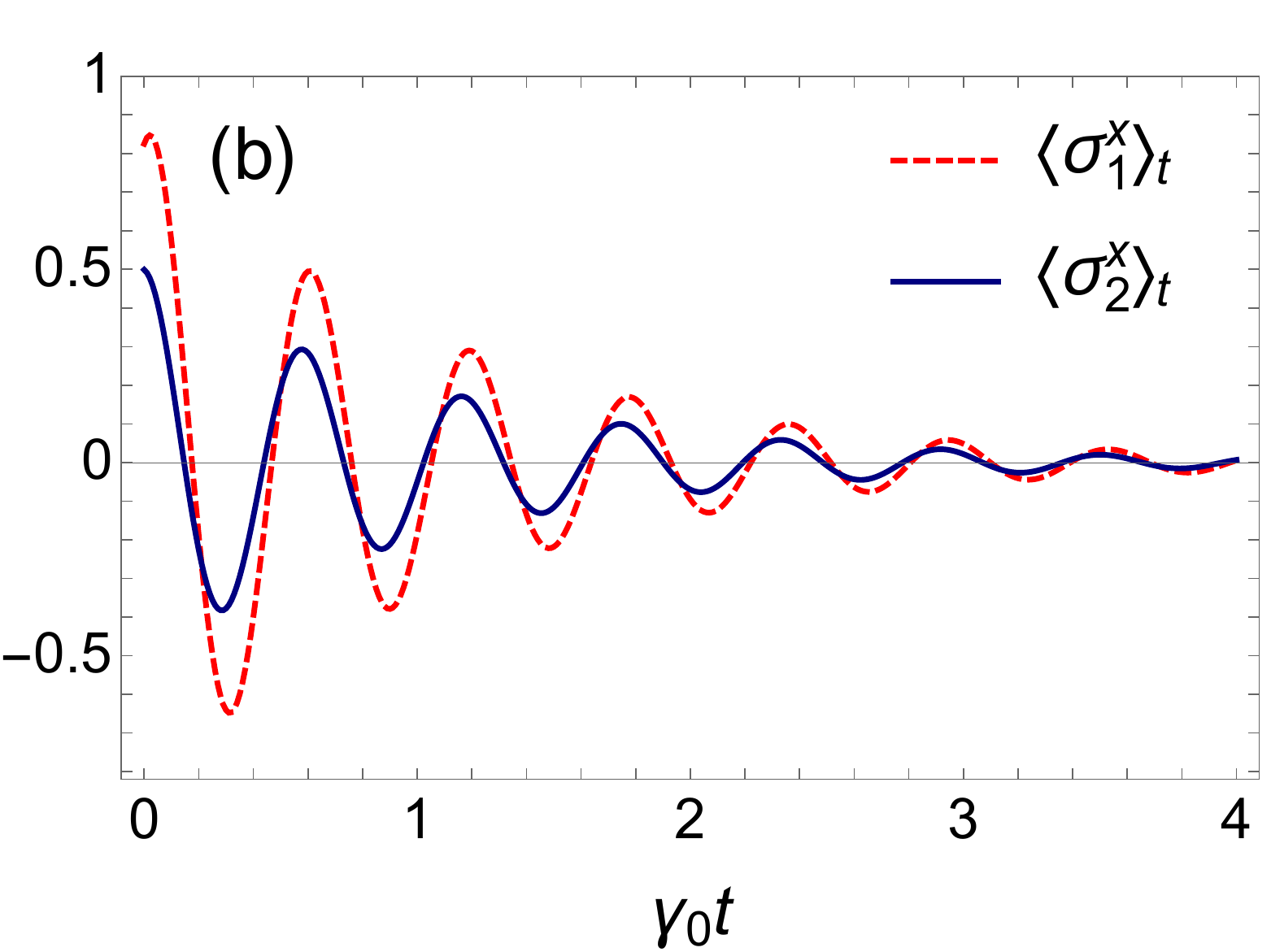}\\
  \caption{Time evolution of $\langle \sigma^x_1 \rangle_t$ (dashed line) and
$\langle \sigma^x_2 \rangle_t$ (solid line). Time is scaled with $\gamma_0$.
The initial state is (a)
$\frac{1}{\sqrt{2}}\left(\ket{G}+\ket{A_R}\right)$ and (b)
 $\frac{1}{\sqrt{2}}\left(\ket{G}+\ket{S_R}\right)$. The parameters have the
same values as in Fig. \ref{FigSync1}.}\label{FigSync2}
\end{figure}

As discussed at the end of Sec. \ref{dmev}, there are particular choices of
the initial state for which only a frequency is involved in the dynamics giving rise to
synchronous dynamics from the initial time, as we show in
Fig. \ref{FigSync2}. Figures \ref{FigSync2}(a) and  \ref{FigSync2}(b)
refer,
respectively, to $\frac{1}{\sqrt{2}}\left(\ket{G}+\ket{A_R}\right)$ and to
$\frac{1}{\sqrt{2}}\left(\ket{G}+\ket{S_R}\right)$.
In both cases, by using Eqs. (\ref{EigensystemL0}) , (\ref{EigensystemLX1_b}),
and
(\ref{MediaA}) it is easy to see that  only one term is present in the averages
of $\langle \sigma^x_1 \rangle_t$  and $\langle \sigma^x_2 \rangle_t$. In Fig.
\ref{FigSync2}(a) the only frequency present is related to  $\Ket{\tau^b_4}$,
while in  Fig. \ref{FigSync2}(b) only $\Ket{\tau^b_3}$ contributes.  The two
signals are synchronized from the beginning,  their
decay rates and the relative phase between the two averages (time delay) being different.
In particular, starting from the initial state
$\frac{1}{\sqrt{2}}\left(\ket{G}+\ket{A_R}\right)$, the two averages end up 
being equal to
\begin{eqnarray}\label{SincroA}
\langle \sigma^x_1 \rangle_t&=&
\frac{|\alpha_A|\,  \mathrm{e}^{- \left( \gamma_0 -V_{\mathrm{r}}
\right) t/2} }{\sqrt{1+|\alpha_A|^2}} \, \cos \left[\nu_S t-
\mathrm{arg}(\alpha_A) \right]
,
 \nonumber \\
\langle \sigma^x_2 \rangle_t&=&
\frac{\mathrm{e}^{- \left( \gamma_0 -V_{\mathrm{r}} \right) t/2}
}{\sqrt{1+|\alpha_A|^2}} \,    \cos \left[\nu_S t \right].
\end{eqnarray}
The relative delay of the two synchronized signals is then equal to
$\mathrm{arg}(\alpha_A)$ [in Fig. \ref{FigSync2}(a) this is $\approx 162^\circ$].

Analogously, when the  the initial state is
$\frac{1}{\sqrt{2}}\left(\ket{G}+\ket{S_R}\right)$, the two averages are given
by
 \begin{eqnarray}\label{SincroS}
\langle \sigma^x_1 \rangle_t&=&
\frac{|\alpha_S|\, \mathrm{e}^{-\left(  \gamma_0  +V_{\mathrm{r}}
\right) t/2} }{\sqrt{1+|\alpha_S|^2}} \,  \,  \nonumber  \\  && \times
   \cos \left[\left(\omega_0 + \frac{1}{2}V_{\mathrm{i}} \right)t -
\mathrm{arg}(\alpha_S) \right]
,
 \nonumber \\
\langle \sigma^x_2 \rangle_t&=&
\frac{\mathrm{e}^{- \left( \gamma_0 - V_{\mathrm{r}} \right)
t/2}}{\sqrt{1+|\alpha_S|^2}} \,
   \cos \left[\left(\omega_0 + \frac{1}{2}V_{\mathrm{i}} \right)t \right].
\end{eqnarray}
The relative delay of the two averages is now equal to
$\mathrm{arg}(\alpha_S)$ [in Fig. \ref{FigSync2}(b), this is $\approx
18^\circ$].

The value of the coefficient   $\gamma_{12}$ can be taken as a
measure of the transition between the common-bath  ($  \gamma_{12}=\sqrt{\gamma_1 \gamma_2}$)
and the separate-bath ($\gamma_{12}=0$) cases. The transition between these two cases is discussed in
Ref. \cite{galve}. In the limit of completely independent atoms, we also have $s_{12}=0$.
In the case of independent atoms ($\gamma_{12}=0$ and $s_{12}=0$), $\langle \sigma^x_1 \rangle_t$ and $\langle \sigma^x_2 \rangle_t$
contain just one term decaying with rates respectively equal to $\gamma_1$ and
$\gamma_2$ and oscillating with frequencies equal  to $\omega_1$ and $\omega_2$.
No synchronization is then possible in this limit case.

In our model, it is the coupling with a common environment that is responsible for the terms in \eqref{MasterEquation} depending on $\gamma_{12}$ and $s_{12}$, causing the spontaneous synchronization described in this section.  However, the presence of a common environment is not the only way of achieving synchronization. In fact, from \eqref{Stilde e Atilde} one can deduce that the roles played by  $\gamma_{12}$ and $s_{12}$ are in some sense complementary. A significant direct interaction between the two atoms (present also in the absence of the environment) can compensate for the absence of a common damping $\gamma_{12}$. This kind of mechanism leading to synchronization was extensively
 discussed in Ref. \cite{syncprobe}, in the presence of direct coupling between the two atoms and of a local bath, where it was also shown that in such a scenario the absence of detuning can be detrimental. In fact, from \eqref{Stilde e Atilde} it follows that for $\gamma_{12}=0 $ and $\delta=0$, a threshold behavior  completely analogous to the one observed in Fig. \ref{FigSP} would occur. Indeed, for any $s_{12}\ge(\gamma_{1}-\gamma_{2})/4 $, $\kappa_S$ is exactly equal to zero, and dynamical synchronization cannot occur. As a matter of fact, in the case of two identical baths and in the absence of detuning, the two atoms are identical, and their trajectories can only be different because of the initial condition.

\section{Super- and subradiance} \label{subr}

In order to study the occurrence of super- and/or subradiance we compute the
evolution of the total radiation rate (given in photons per second)
\cite{Haroche}
\begin{equation}\label{RadiationRate}
  I(t)= \sum_{i,j} \gamma_{ij} \langle \sigma_i^+ \sigma_j^- \rangle_t,
\end{equation}
given by the average of the operator $Q=\sum_{i,j} \gamma_{ij}  \sigma_i^+
\sigma_j^- $.
All the matrix elements associated with the operators $\langle \sigma_i^+
\sigma_j^-\rangle $ exist in the Liouvillian subspace  $\mathcal{L}^a$.
It follows that $I(t)$ is given by
\begin{equation}\label{AverageI}
  I(t)= \sum_{i} p_{0\, i}^{a}\,   \langle \tau^a_i  \rangle_{Q}   \,
\mathrm{e}^{\lambda_i^a t},
\end{equation}
where $\langle \tau^a_i  \rangle_{Q}= \mathrm{Tr}(Q \, \tau^a_i)$. In particular,
$\langle \tau^a_1  \rangle_{Q}=0$ [the ground state does not contribute to
$I(t)$]. In the two cases when the initial states are chosen, respectively, as
$\Ket{A_RA_R}$ and $\Ket{S_RS_R}$, by using Eqs. (\ref{rhoAt}) and
(\ref{rhoSt}), one finds
\begin{eqnarray}\label{AverageIas}
I(t)&=& -\lambda_6^a\mathrm{e}^{\lambda_6^a t} =\left(\gamma_0-
V_\mathrm{r}\right)   \mathrm{e}^{-\left(\gamma_0- V_\mathrm{r}\right)
t}  \nonumber  \\ &&\qquad \qquad(\mathrm{for}\,\,\Ket{\rho_0}=\Ket{A_RA_R}),
\nonumber \\
I(t)&=&-\lambda_5^a\mathrm{e}^{\lambda_5^a t} =\left(\gamma_0+
V_\mathrm{r}\right)   \mathrm{e}^{-\left(\gamma_0+ V_\mathrm{r}\right)
t}  \nonumber \\&& \qquad \qquad (\mathrm{for}\,\,\Ket{\rho_0}=\Ket{S_RS_R}).
\end{eqnarray}
\begin{figure}[t]
  \centering
  \includegraphics[width=0.47  \textwidth]{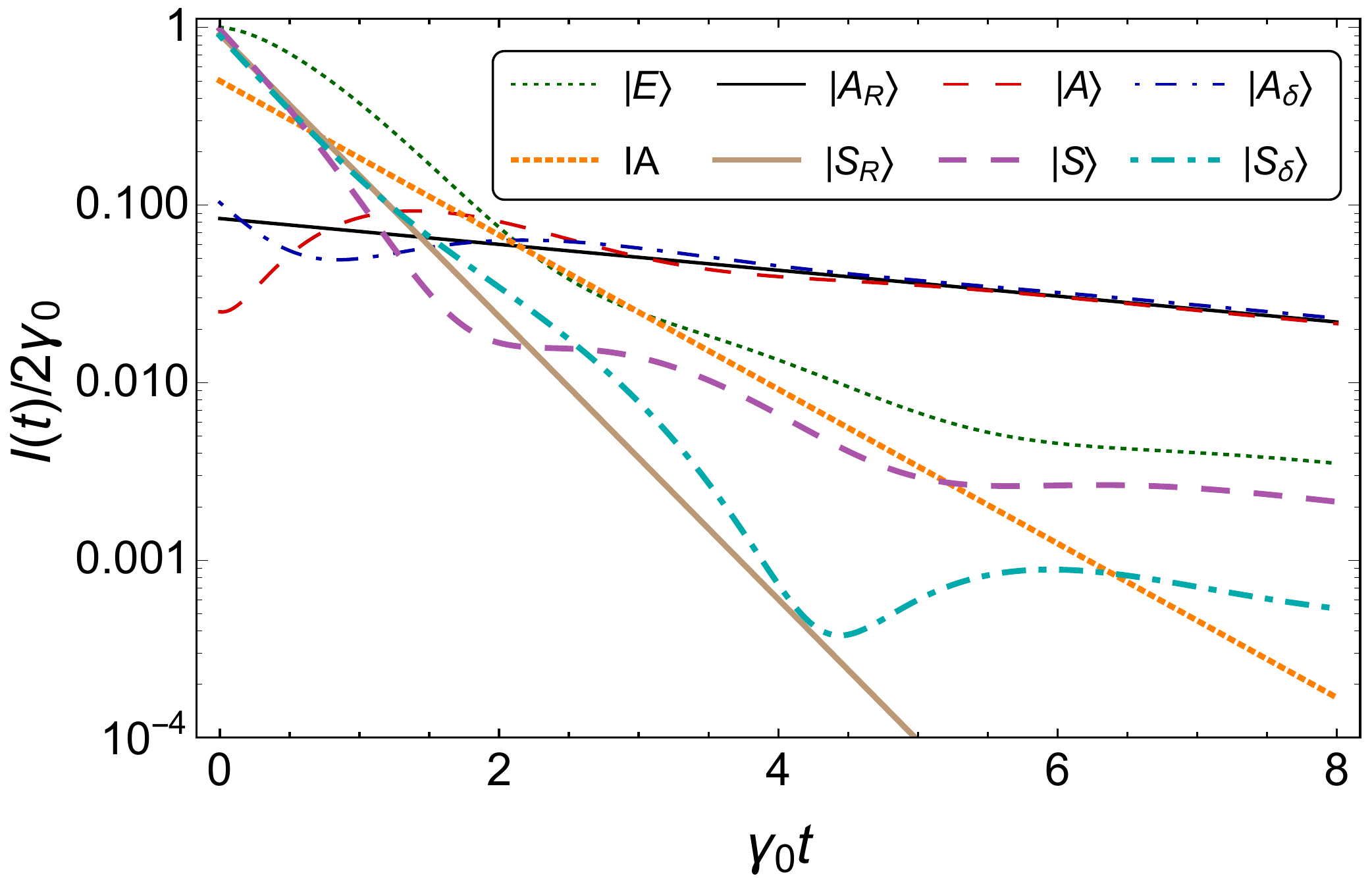}\\
  \caption{Time evolution of $I(t)/2\gamma_0$ (logarithmic scale) for several initial
conditions (see the legend; the curves have different styles and thicknesses). Time is scaled with $\gamma_0$. The parameters have
the same values as in Fig. \ref{FigSync1}. Here $\ket{S}$ and $\ket{A}$ are the
symmetric and antisymmetric states, respectively, and $\ket{S_\delta}$ and $\ket{A_\delta}$ are the
one-excitation states diagonalizing the dressed Hamiltonian $H_S+ H_{LS}$.
The IA case is the one of independent atoms having individual decay rates
equal to $\gamma_0$, starting from an arbitrary one-excitation
state.}\label{FigSR1}
\end{figure}
Equations (\ref{AverageIas}) describe how the two-atom system  emits when initially
prepared either in the state $\ket{A_R}$ or in $\ket{S_R}$. Both emissions are
characterized by only one decay rate $\gamma_0- V_\mathrm{r}$ and
$\gamma_0+ V_\mathrm{r}$ respectively, displaying the known different radiances
for nonvanishing $V_\mathrm{r}$. This feature is a first
quantitative link with the synchronization parameter
$\kappa_S=V_\mathrm{r}$ defined in \eqref{Sincronizzazione}. When
$\kappa_S$ increases towards its maximum given by $\gamma_0$ (faster
synchronization), the decay rates in \eqref{AverageIas} go, respectively, to $0$
(subradiance) and to $\gamma_1+\gamma_2$ (superradiance).

Figure \ref{FigSR1} shows the evolution of $I(t)$ (on a logarithmic scale) for several
initial conditions. One sees that, by starting from $\ket{S_R}$ or $\ket{A_R}$,
the decay of $I(t)$ is governed by only one decay rate [see \eqref{AverageIas}].
The same holds for the case of independent atoms (IAs) $\gamma_{12}=
s_{12}=0$) having individual decay rates equal to $\gamma_0$. For any other
initial states, the evolution of  $I(t)$ is not an exponential function as it
depends on more than one decay rate.
Starting from the symmetric  and antisymmetric states $\ket{S}$ and $\ket{A}$, respectively,
the total radiation rates are initially the larger and the smaller among all the
possible one-excitation states. However, during the dynamics,  part of the
population remains trapped in $\ket{A_R}$, whose decay rate (the decay rate of
the eigenvector $\Ket{\tau^a_6}$) governs the final part of the evolution of
$I(t)$. Similar considerations hold for the initial states $\ket{S_\delta}$ and
$\ket{A_\delta}$ [the states that diagonalize the dressed Hamiltonian $H_S$ +
$H_{LS}$
of \eqref{MasterEquation}]. Also the case in which initially both atoms are
excited, the condition typically considered in the studies of super- and
subradiance, is plotted.

\begin{figure}[t]
  \centering
  \includegraphics[width=0.4 \textwidth]{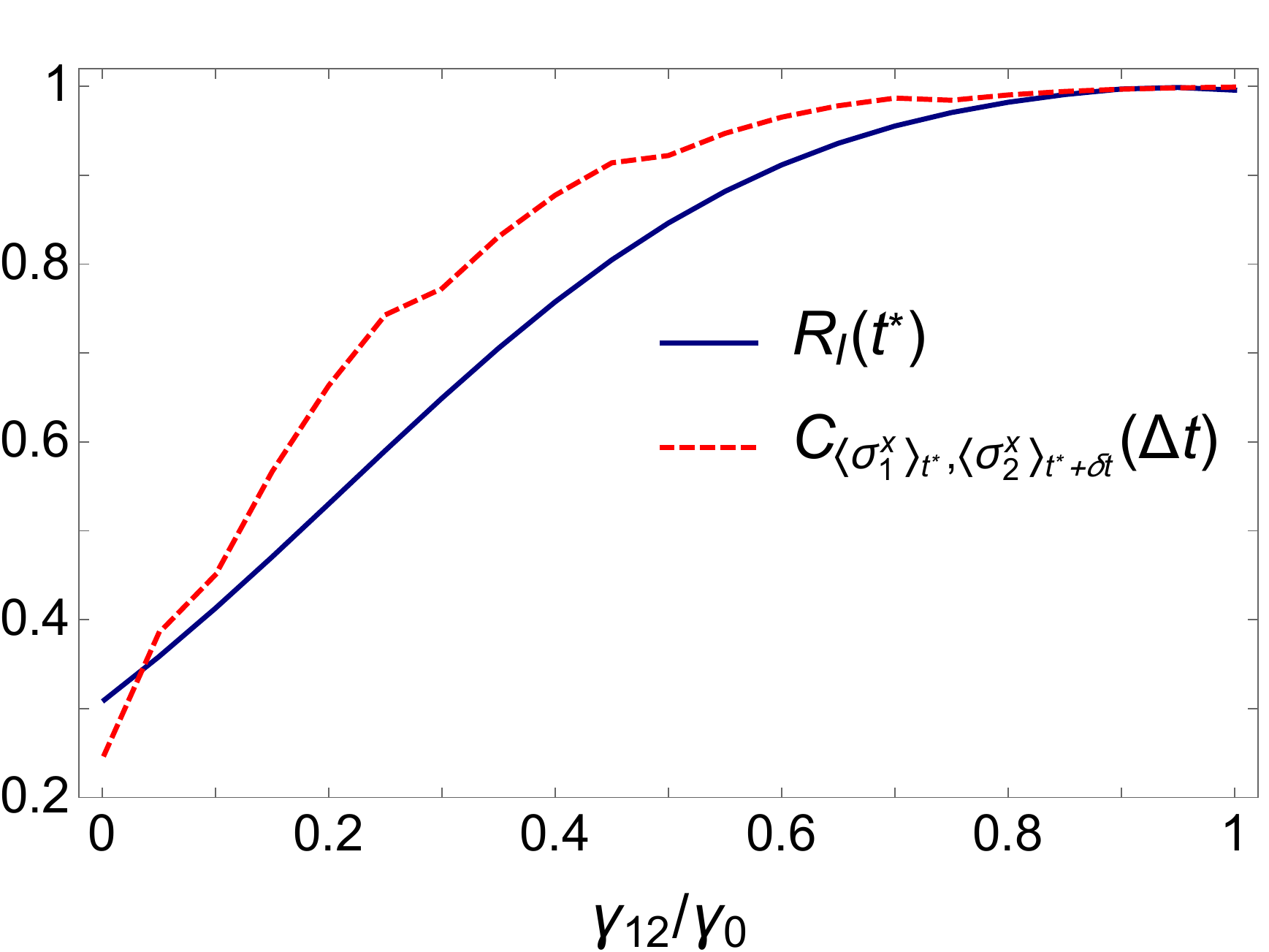}\\
  \caption{Comparison between the synchronization measure
   $C_{\langle
\sigma^x_1 \rangle_{t^*},\langle \sigma^x_2 \rangle_{(t^*+\delta t)}}(\Delta
t)$ $(\Delta t=2/\gamma_0)$ and the ratio $R_I(t^*)$, both computed at
$t^*=5/\gamma_0 $ for the initial state
$\frac{1}{2}(\ket{e}+\ket{g})(\ket{e}+\ket{g})$. The values of the parameters are $\gamma_1 =\gamma_2=\gamma_0$,
$s_{12} =0.6 \gamma_0$, $\delta= \gamma_0$, and $\omega_0=10 \gamma_0$. For
$\gamma_{12}=\gamma_0$, the synchronization parameter is such that  $\kappa_S^{-1}
\gamma_0 \approx 1.2$, while for $\gamma_{12}=0.3 \gamma_0$, $\kappa_S^{-1}  \gamma_0
\approx 4.3$.}\label{FigComparison}
\end{figure}

In general (i.e., for most atoms initial states), the large-time behavior of $I(t) $ is governed by the long-lasting eigenvector:
 $I_{SR}(t)= p_{0\, 6}^{a}\,   \langle \tau^a_6  \rangle_{Q}   \,
\mathrm{e}^{\lambda_6^a t}= - p_{0\, 6}^{a}\lambda_6^a \mathrm{e}^{\lambda_6^a
t} $, which is the subradiant contribution of  $ \Ket{\tau^a_6}$ weighted by the
overlap between this eigenmode and  the considered initial state.
Furthermore, the ratios
between the other contributions and the one with the slowest decay all  decay
faster than $\mathrm{e}^{(\mathrm{Re}[\lambda_3^a]-\lambda_6^a)
t}=\mathrm{e}^{(\mathrm{Re}[\lambda_4^a]-\lambda_6^a) t}=\mathrm{e}^{-
V_\mathrm{r} t}$, that is, with a characteristic time that is equal to the
synchronization time [see \eqref{Sincronizzazione}], $\kappa_S^{-1}=1/
V_\mathrm{r}$. This clearly shows how the time scales of  synchronization
and of subradiant emission are intimately related.

This analysis establishes analytically that the two phenomena under study, that
is, synchronization and super- and subradiance, occur at the same time scale.
This is displayed in Fig.  \ref{FigComparison}, representing the delayed synchronization measure
$C(\Delta t)$ defined in Eq. (\ref{eq:pears}) and a subradiance
parameter $R_I(t)$, defined as the minimum between $I(t)/I_{SR}(t)$ and its
inverse $I_{SR}(t)/I(t)$. The parameter $R_I(t)$ is such that it goes to 1 when $I(t)$
approaches its large time value $I_6(t)$ (from above or below). $C(\Delta t)$
 is computed using in   \eqref{eq:pears} $\langle \sigma^x_i \rangle_t$ as
local operators and allowing for a time delay
between the two signals. The comparison is done by varying $\gamma_{12}$ from
$0$ to its maximum $\gamma_0$ ($\gamma_1=\gamma_2=\gamma_0$ in the
figure). The above parameters are computed at a time $t^*=5/\gamma_0$ such that
for low  values of $\gamma_{12}$, $\langle \sigma^x_1 \rangle_t$ and $\langle
\sigma^x_2 \rangle_t$ are not yet synchronized and  $I(t)$ contains still more
than one
contribution from the eigenstates of $\mathcal{L}^a$. By increasing $\gamma_{12}$,
 $V_\mathrm{r}$ increases and we may observe at the chosen time $t^*$ the
occurrence of synchronization and $I(t)$ approaching its subradiant
contribution $I_{SR}(t)$.

\section{Dephasing noise}\label{depnoise}
So far we have discussed the case of a fully dissipating environment and
found a perfect connection between the time scales of emergence of
synchronization and  superradiance. As discussed in Ref. \cite{giorgi}, the
presence of decohering noise even if into a common environment does not favor synchronization.
Here, we also show that it alters the interplay with superradiance.

Although the presence of decoherence can be treated exactly \cite{pse}, in the
weak-coupling limit dephasing effects can be included by adding  a
phenomenological term to the master equation of \eqref{MasterEquation}
\cite{ban}:
\begin{equation}
\mathcal{D}_{{\rm dep}}= \sum_{ij}  \gamma_{ij}^d \left( \sigma_i^z \rho  \sigma_j^z
-\frac{1}{2} \left\{ \sigma_j^z \sigma_i^z,\rho \right\} \right),
\end{equation}
where $\gamma_{21}^d =\gamma_{12}^d$ is assumed to be real and $\gamma_{12}^d
\leq \sqrt{\gamma_{11}^d\gamma_{22}^d}$.
One can verify that this extra term does not modify the decomposition into the
direct sum of sectors of the new Liouvillian.  In particular, looking at $
\mathcal{L}_b$, its new eigenvalues $\lambda_i^b$ are all equal to the old ones
apart from the common term $2(\gamma_{11}^d+\gamma_{22}^d)$ that must be
subtracted from all of them.
 It follows that, analogously to what was found in Sec.  \ref{Sec: synchronization},
the time scales for the emergence of quantum synchronization (survival of one
term in the average of arbitrary single-atom operators, with respect to the
other contributions) remain the same, given by  $\kappa_S^{-1}=1/
V_\mathrm{r}$. However, the amplitude of this survival term now is lowered
at any time $t$ by a factor $\mathrm{e}^{-2(\gamma_{11}^d+\gamma_{22}^d)t}$.

As for $ \mathcal{L}_a$, the situation is different and we can distinguish two
main cases. In the case when $\gamma_{11}^d+\gamma_{22}^d= 2\gamma_{12}^d$, $
\mathcal{L}_a$ is not modified and our previous analysis remains the same. When
this condition is not satisfied, in the general case we do not  have  analytical
expressions for the eigenvalues  $\mathcal{L}_a$ anymore. However, a numerical
analysis shows that the ratio between the  exponentials governed by the two
smallest (in modulus) decay rates  now depends  on  the value of
$\gamma_{11}^d+\gamma_{22}^d- 2\gamma_{12}^d$. This means that in general the
time scales of spontaneous synchronization and of the emergence of subradiant
emission are not  equal any more.

\section{Conclusions}\label{conclusions}

Superradiance is a genuinely collective phenomenon, as it cannot be
explained by taking only individual constituents of the whole atomic system. The
main question addressed in this paper  concerns the possible evidence  of such a global
effect on the dynamics and synchronization of local observables. In fact, we have found a strict analytical
relationship between subradiance and the dynamical synchronization of local
trajectories. Considering two detuned qubits coupled to a dissipative bath, by
means of the Liouville formalism,  we have studied the microscopic mechanisms
responsible for the two phenomena and found that, actually, synchronization and
super- and subradiance are closely related in our model. As a matter of fact,
the synchronization between averages of arbitrary single atom operators is due
to the presence of longstanding coherences between the ground state and a state
that can be identified as responsible for subradiant emission, whose experimental detection is known to be more demanding \cite{devoe,subradiance}. The fact
that the decay rate due to such a subradiant state is twice the decay rate of
the coherence responsible for synchronization clearly shows that the two phenomena
are intimately related  and occur on the same time scale. Finally, we have shown that  by adding the effects of a dephasing noise, the correspondence between synchronization and collective
super- and subradiance effects can be altered.

The case of two atoms we have  discussed here offers the
advantage that a fully analytical treatment is available that makes patent the
liaison between subradiance and synchronization. The comparison in
the case of an ensemble of $N$
atoms cannot be treated analytically and  requires numerical or approximated descriptions, like the mean field adopted, for instance,  in Ref. \cite{hollande2014}
to assess the steady state in the presence of driving. However, in the case of two interacting clouds of different atomic  species \cite{Haroche}, in some symmetric configurations,  one can expect that  results similar to the ones obtained here could be valid with the occurrence of synchronization between macroscopic quantities such as the sum of individual spins in each cloud.  Indeed, the time evolution of the beating note  for sufficient long times  should provide a signature of such a phenomenon.  On the other hand, the study of the effects of the driving cannot be fully analytical even in the case of only two atoms. In this case open questions remain about the connection between subradiance and synchronization and their relative time scales.

The occurrence of synchronization  in  classical systems, as well as its definition in
a broad spectrum of regimes, is well established \cite{Pikovsky}.
When it comes to the quantum realm, it becomes interesting to identify
specific quantum signatures of this effect. Recently, there
have been many attempts to quantify the appearance of synchronization also through quantum
indicators of (global) correlations \cite{chapter}.  Our work represents an original contribution
also in this context  given that it offers an explanation of spontaneous synchronization of local observables as the
manifestation of collective quantum interference. In other words, the present analytical description through local generic observables dynamics allows us to establish
how synchronization witnesses a collective quantum phenomenon of subradiance,
representing then a genuine quantum feature of synchronization, intimately different from its
classical counterpart.

\begin{acknowledgments}
B.B. thanks D. Viennot for useful discussions. Fundings from EU project QuProCS (Grant Agreement No. 641277), MINECO, AEI/FEDER (Grants No. FIS2014-60343-P and No. FIS2016-78010-P), ``Vicerectorat d'Investigaci\'o
 i Postgrau'' through the visiting professors program of the UIB and from ``Soutien aux enseignants-chercheurs et chercheurs -
Nouveaux arrivants 2016'' of the UFC are acknowledged.
\end{acknowledgments}

\appendix

\section{Liouville representation} \label{AppendixL}
To treat the dynamics of the system, we adopted in Sec. \ref{dmev} the so-called Liouville
representation of the density matrix, defined by
\begin{equation}
\rho = \sum_{i,j=1}^4 \rho_{ij} |i\rangle \langle j| \xrightarrow{HS} \Ket{\rho}
= \sum_{i,j=1}^4 \rho_{ij}  \Ket{ij } ,
\end{equation}
where $ \Ket{ij }= |i\rangle \otimes |j \rangle$. The inner product in the
Hilbert-Schmidt space is defined by $\llangle \tau  \Ket{ \rho } =
\tr(\tau^\dagger \rho)$. It follows that  for any operator of the system $O$ 
we have
\begin{equation}
\Ket{ O \rho }  =  O \otimes \mathbb{I} \Ket{\rho }, \quad
\Ket{\rho O }  =  \mathbb{I} \otimes O^\transp \Ket{\rho}
\end{equation}
where $O^\transp$ denotes the matrix transposition of $O$ and $\mathbb{I}$ the
four-dimensional identity matrix. By using these rules, we obtain the Liouville
representation of the master equation of \eqref{MasterEquation}, $|\dot \rho
\rrangle = \mathcal{L} |\rho \rrangle$,
where  $ \mathcal{L}$ is defined by
\begin{eqnarray}\label{CostruzioneL}
& & \mathcal{L}  =  -i \left(H_\mathrm{eff} \otimes \mathbb{I} - \mathbb{I}
\otimes H_\mathrm{eff}^\transp \right)   \\ &&+ \frac{1}{2} \sum_{ij}
\gamma_{ij} \left[2  \sigma_i^- \otimes  (\sigma_j^+)^\transp - \sigma_j^+
\sigma_i^-  \otimes \mathbb{I} -\mathbb{I}  \otimes  (\sigma_j^+  \sigma_i^-
)^\transp \right], \nonumber
\end{eqnarray}
where $H_{\mathrm{eff}}=H_S+H_{LS}$.
The Liouville representation of the master equation is thus a Schr\"{o}dinger
equation governed by a non-self-adjoint generator $ \mathcal{L}^\dagger \neq
\mathcal{L}$.

\subsection{Decomposition of $\mathcal{L}$} \label{Ldecomp}

By using \eqref{CostruzioneL}, one  finds that the total Liouvillian can be
decomposed in five Jordan blocks.
The Hilbert-Schmidt space $\mathcal H = \mathbb C^{16}$ can be thus decomposed
into five independent subspaces: $\mathcal H = \bigoplus_{\mu} \mathcal H_\mu$,
$\mu \in \{a, b, c, d, e\}$, with $\dim \mathcal H_a = 6$, $\dim \mathcal H_{b}
= \dim \mathcal H_{c}=4$, and $\dim \mathcal H_{d} =\dim \mathcal H_{e} = 1$,
such that $  \mathcal{L} = \bigoplus_{\mu}  \mathcal{L}_\mu$ ($ \mathcal{L}_\mu$
being an operator of $\mathcal H_\mu$). Subspace $\mathcal H_a$ is generated by the basis elements
$\ket{ee}\bra{ee}$, $\ket{eg}\bra{eg}$, $\ket{eg}\bra{ ge}$, $\ket{ge}\bra{eg}$,
$\ket{ge}\bra{ge}$, and $\ket{gg}\bra{gg}$; $\mathcal H_b$  by $\ket{ee}\bra{eg}$,
$\ket{ee}\bra{ge}$, $\ket{eg}\bra{ gg}$, and $\ket{ge}\bra{gg}$; $\mathcal H_c$  by
their complex conjugate counterparts $\ket{eg}\bra{ee}$, $\ket{ge}\bra{ee}$,
$\ket{gg}\bra{ eg}$, and $\ket{gg}\bra{ge}$; $\mathcal H_d$ by $\ket{ee}\bra{gg}$;
and $\mathcal H_e$ by its conjugate $\ket{gg}\bra{ee}$.

The matrix representations of $\mathcal{L}_\mu$ in the basis of $\mathcal
H_{\mu}$ are given by
\begin{widetext}\center
\begin{eqnarray}\label{L0}
 \mathcal{L}_a&=& \left(
\begin{array}{cccccc}
 - \left(\gamma_1+\gamma_2\right) & 0 & 0 & 0 & 0 & 0 \\
 \gamma_2 & - \gamma_1 &-\frac{\gamma_{12}}{2}+  i s_{12} &
-\frac{\gamma_{12}}{2}-  i s_{12} & 0 & 0 \\
  \gamma_{12} & -\frac{\gamma_{12}}{2}+  i s_{12} &-\frac{\gamma_1+\gamma_2}{2}
- i \delta  & 0 & -\frac{\gamma_{12}}{2}-  i s_{12} & 0 \\
  \gamma_{12} & -\frac{\gamma_{12}}{2}-  i s_{12} & 0 & -
\frac{\gamma_1+\gamma_2}{2}+  i \delta & -\frac{\gamma_{12}}{2}+  i s_{12}& 0 \\
  \gamma_1 & 0 & -\frac{\gamma_{12}}{2}-  i s_{12} & -\frac{\gamma_{12}}{2}+  i
s_{12} & - \gamma_2 & 0 \\
 0 &  \gamma_1 &  \gamma_{12} & \gamma_{12} &  \gamma_2 & 0 \\
\end{array}
\right), \end{eqnarray} \begin{eqnarray}\label{LX1}
\mathcal{L}_b&=& \left(
\begin{array}{cccc}
 -\gamma_1-\frac{\gamma_2}{2}-i \left(\omega_0-\frac{\delta}{2} \right) &
-\frac{\gamma_{12}}{2}+i s_{12} & 0 & 0 \\
 -\frac{\gamma_{12}}{2}+i s_{12} & -\frac{\gamma_1}{2}- \gamma_2-i \left(\omega
_0+\frac{\delta}{2} \right) & 0 & 0 \\
  \gamma_{12} &  \gamma_2 &- \frac{\gamma_1}{2} -i \left(
\omega_0+\frac{\delta}{2} \right) & -\frac{\gamma_{12}}{2}-i s_{12} \\
  \gamma_1 &  \gamma _{12} & -\frac{\gamma_{12}}{2}-i s_{12} & -\frac{\gamma
_2}{2} - i \left(\omega _0 -\frac{\delta}{2} \right) \\
\end{array}
\right),
\end{eqnarray}
\end{widetext}
while $\mathcal{L}_c$ is the conjugate of $\mathcal{L}_b$, $\mathcal{L}_d$ has just one
element equal to  $-\gamma_0 - 2 i \omega_0$, and $\mathcal{L}_e$ is the
conjugate of $\mathcal{L}_d$.

\subsection{Eigenvectors and eigenvalues} \label{Eigenvectors and eigenvalues}

Let $\{\lambda_i^\mu\}$, $\{\tau_i^\mu\}$,  and $\{\bar{\tau}_i^{\mu}\}$ be the
eigenvalues and the right and  left instantaneous eigenvectors of
$\mathcal{L}$. They are defined by
\begin{eqnarray}\label{eig}
\mathcal{L} \Ket{\tau_i^\mu} & = & \lambda_i^\mu\Ket{\tau_i^\mu }, \quad
\mathcal{L}^\dagger \Ket{\bar{\tau}_i^{\mu}}  =  \lambda_i^{\mu\,*}
\Ket{\bar{\tau}_i^{\mu } }, \quad \mathrm{being} \nonumber \\
\llangle \bar{\tau}_i^{\mu } \Ket{\tau_j^{\nu } } &=& \tr(\bar{\tau}_i^{\mu
\,\dagger} \tau_j^{\nu }) =\llangle \bar{\tau}_i^{\mu } \Ket{\tau_i^{\mu} }
\delta_{\mu \nu}\delta_{ij}.
\end{eqnarray}
We can write the identity in the  Hilbert-Schmidt space, $\mathcal{I}$, by means
of the right and  left instantaneous eigenvectors of $\mathcal{L}$ as
\begin{equation}\label{identity}
\mathcal{I}= \bigoplus_{\mu}  \sum_{i}\frac{\Ket{\tau_i^\mu}    \llangle
\bar{\tau}_i^{\mu }  |}{ \llangle \bar{\tau}_i^{\mu } \Ket{\tau_i^\mu}    }.
\end{equation}

\subsubsection{ $\mathcal{L}_a$}

The eigenvalues and the right eigenvectors of $\mathcal{L}_a$ are
\begin{eqnarray}\label{EigensystemL0}
\begin{cases}
   \Ket{\tau^a_1} = \Ket{G G} ,   & \lambda_1^a=0, \\
   \Ket{\tau^a_2} = \Ket{E E} -x_1 \Ket{ S_R S_R }-x_2 &\!\!\!\! \!\! \Ket{
A_RA_R }  -z \Ket{ S_R A_R }  \\ \qquad \quad -z^* \Ket{ A_R S_R }
   +y \Ket{G G}  ,      & \lambda_2^a=- 2 \gamma_0, \\
   \Ket{\tau^a_3} =  \frac{\Ket{ S_R A_R }}{\langle A_R \ket{S_R}} - \Ket{G  G}
,   & \lambda_3^a=- \gamma_0- i\,V_\mathrm{i}, \\
  \Ket{\tau^a_4} = \frac{\Ket{ A_R S_R }}{\langle S_R \ket{A_R}} - \Ket{G  G},
   & \lambda_4^a=- \gamma_0+ i V_\mathrm{i}, \\
     \Ket{\tau^a_5} = \Ket{ S_R S_R } - \Ket{G  G} ,  \quad    & \lambda_5^a=-
\gamma_0- V_\mathrm{r}, \\
         \Ket{\tau^a_6} = \Ket{ A_R A_R} - \Ket{ G  G } ,  \quad    &
\lambda_6^a=- \gamma_0+ V_\mathrm{r} ,\\
 \end{cases}
\end{eqnarray}

\noindent where  $\Ket{ S_R S_R}$ and $\Ket{ A_R A_R}$ are the projectors of
the kets $\ket{A_R}$ and $\ket{S_R}$, defined in \eqref{Stilde e Atilde}, and
where $x_1$, $x_2$, $y$, and $z$ have a complicated dependence on the various parameters of the model, not reported here.

We remark that the eigenvector $\Ket{\tau^a_1}=\Ket{GG}$ associated with the
eigenvalue $\lambda_1^a= 0$ is the steady state of the two-atom system.

The left eigenvectors in the sector $\mathcal{L}_a$ are (we report the right
eigenvectors of $\mathcal{L}_a^\dagger$)
\begin{eqnarray}\label{LeftL0}
\begin{cases}
   \Ket{\bar{\tau}^a_1} = \Ket{\mathbb{I} } ,  \\
   \Ket{\bar{\tau}^a_2} = \Ket{E E} ,  \\
   \Ket{\bar{\tau}^a_3} =  \frac{\Ket{ S_L A_L }}{\mathrm{Tr}(\ket{S_L}
\bra{A_L})} + x \Ket{E  E} , \\
  \Ket{\bar{\tau}^a_4} = \frac{\Ket{ A_L S_L }}{\mathrm{Tr}(\ket{A_L}
\bra{S_L})} +x^* \Ket{E  E},     \\
     \Ket{\bar{\tau}^a_5} = \Ket{ S_LS_L } +y_1 \Ket{E  E} ,   \\
         \Ket{\bar{\tau}^a_6} = \Ket{ A_L A_L} +y_2 \Ket{ E  E } , \\
 \end{cases}
\end{eqnarray}
where $\Ket{\mathbb{I} }$ is the vector associated with the identity in the 
four-dimensional space (in the decoupled basis it is $\Ket{\mathbb{I} }
=\Ket{EE}+\Ket{eg\, eg }+\Ket{ge\, ge}+\Ket{GG} $) and where
\begin{eqnarray}\label{coefficientsleft}
  y_1 &=&
\frac{\gamma_0+V_\mathrm{r}+(\gamma_1-\gamma_2)\frac{1-|\alpha_S|^2}{
1+|\alpha_S|^2}}{\gamma_0-V_{\mathrm{r}}} , \nonumber \\
  y_2 &=&
\frac{\gamma_0-V_\mathrm{r}+(\gamma_1-\gamma_2)\frac{1-|\alpha_A|^2}{
1+|\alpha_A|^2}}{\gamma_0+V_{\mathrm{r}}} ,  \\
  x&=& |\alpha_A| \frac{\gamma_1  +\gamma_2 \alpha_A
\alpha_S^*+\gamma_{12}(\alpha_A+\alpha_S^*)}{(1+|\alpha_A|^2) \langle A_L
\ket{S_L}  (\gamma_0 +  i V_{\mathrm{i}})}.\nonumber
\end{eqnarray}

\subsubsection{$\mathcal{L}_b$, $\mathcal{L}_c$, $\mathcal{L}_d$ and
$\mathcal{L}_c$}

The eigenvalues of  $\mathcal{L}_b$ are
\begin{eqnarray}\label{EigensystemLX1}
\begin{cases}
   \lambda_1^b=-\frac{1}{2} \left[3\gamma_0
+V^*\right]-i \omega_0,  \\  \lambda_2^b=-\frac{1}{2} \left[ 3\gamma_0
-V^* \right]-i \omega_0,  \\
  \lambda_3^b=-\frac{1}{2} \left(
\gamma_0+V\right)-i \omega_0, \\\lambda_4^b=-\frac{1}{2} \left(\gamma_0
-V \right)-i \omega_0  .
 \end{cases}
\end{eqnarray}
Concerning the right  the left eigenvectors of  $\mathcal{L}_b$, we found the analytical expressions  for  $\Ket{\tau^b_{3,4}}$ and $\Ket{\bar{\tau}^b_{1,2}}$  (we report the right
eigenvectors of $\mathcal{L}_b^\dagger$):
\begin{eqnarray}\label{EigensystemLX1_b}
   \Ket{\tau^b_3} =  \Ket{S_R G},   \qquad
  \Ket{\tau^b_4} = \Ket{A_R G},
\end{eqnarray}
and
\begin{eqnarray}\label{LeftLX1}
   \Ket{\bar{\tau}^b_1} = \Ket{E S_L},  \qquad
   \Ket{\bar{\tau}^b_2} = \Ket{E A_L} .
\end{eqnarray}
The expressions for the other eigenvectors are particularly cumbersome and are then not reported here.

The eigenvalues in the sector $\mathcal{L}_c$ are the conjugates of the ones in
$\mathcal{L}_b$, while the right and left  eigenvectors are given by
$\tau^c_i=\tau^{b\,\dagger}_i$ and $\bar{\tau}^c_i=\bar{\tau}^{b\,\dagger}_i$.

The only eigenvalue in  $\mathcal{L}_d$ is equal to $\lambda_1^d=-\gamma_0- 2 i
\omega_0$ and the one in  $\mathcal{L}_e$ is $\lambda_1^e=\lambda_1^{d\,*}$. The
corresponding right and left eigenvectors are $ \Ket{\tau^d_1} =
\Ket{\bar{\tau}^d_1} =\Ket{E G} $ and $ \Ket{\tau^e_1} = \Ket{\bar{\tau}^e_1}
=\Ket{G E} $.

\subsection{Rate equations}

Here we report the equations of motion  for the density-matrix elements
belonging to the sector $\mathcal{L}_a$. In the decoupled basis the rate
equations for the populations and the coherences between $\ket{eg}$ and
$\ket{ge} $  can be immediately derived using the form of $\mathcal{L}_a$ of
\eqref{L0}  (we use the notation  $\bra{I} \rho \ket{J}=\rho_{I,J}$).
In this basis, the populations of $\ket{eg}$ and $\ket{ge} $ are coupled with
their coherences.
 The special role played by the states $ \ket{S_L}$ and $ \ket{A_L}$ emerges
clearly when we derive the equations of motion for their populations, together
with the equations of motion  for their coherences, and for the ground state.
These can be obtained by recombining the rate equations in the decoupled basis,
exploiting \eqref{Stilde e Atilde}, or, equivalently, by using the left
eigenvectors of \eqref{LeftL0} and the relation, for an arbitrary operator $O$,
$ \frac{\mathrm{d}  \langle O \rangle  }{\mathrm{d}t}=[\mathrm{Tr} (\rho
\mathcal{L}^\dagger O^\dagger)]^* $:
\begin{eqnarray}
\frac{\mathrm{d}  \rho_{E,E}  }{\mathrm{d}t}& = &-\left(\gamma_1+\gamma_2\right)
 \rho_{E,E} , \nonumber 
      \end{eqnarray}
\begin{eqnarray}  \label{rate equation}
\frac{\mathrm{d}  \rho_{S_L,S_L}  }{\mathrm{d}t} &=
&\left[\gamma_0+V_\mathrm{r}-(\gamma_1-\gamma_2)\frac{1-|\alpha_A|^2}{
1+|\alpha_A|^2}\right] \rho_{E,E}  \nonumber  \\
  &&   - \left( \gamma_0  +V_{\mathrm{r}}\right)  \rho_{S_L,S_L}  ,
 \nonumber 
\\
 \frac{\mathrm{d}  \rho_{S_L,A_L  }}{\mathrm{d}t} &= & |\alpha_A| \frac{\gamma_1
+\gamma_2 \alpha_S \alpha_A^*+\gamma_{12}(\alpha_S+\alpha_A^*)}{(1+|\alpha_A|^2)
}  \rho_{E,E}  \nonumber  \\
  &&   - \left( \gamma_0  + i V_{\mathrm{i}}\right)  \rho_{S_L,A_L} ,
    \nonumber
\\
\frac{\mathrm{d}  \rho_{A_L,S_L  }}{\mathrm{d}t} &= & |\alpha_A| \frac{\gamma_1
+\gamma_2 \alpha_A \alpha_S^*+\gamma_{12}(\alpha_A+\alpha_S^*)}{(1+|\alpha_A|^2)
}  \rho_{E,E}  \nonumber  \\
  &&   - \left( \gamma_0  - i V_{\mathrm{i}}\right)  \rho_{A_L,S_L} ,
    \nonumber  \\
     \frac{\mathrm{d}  \rho_{A_L,A_L  }}{\mathrm{d}t} &= &
\left[\gamma_0-V_\mathrm{r}+(\gamma_1-\gamma_2)\frac{1-|\alpha_A|^2}{
1+|\alpha_A|^2}\right] \rho_{E,E}  \nonumber  \\
  &&   - \left( \gamma_0  -V_{\mathrm{r}}\right)  \rho_{A_L,A_L}  , \nonumber \\  \nonumber
  \frac{\mathrm{d  \rho_{G,G}  }}{\mathrm{d}t}&
=&\frac{1+|\alpha_A|^2}{|1+\alpha_A^2|^2} \Big[  (1+|\alpha_A|^2)
\left(\gamma_0+V_\mathrm{r}     \right)\nonumber  \\
 && \!\!\!\!\!\! \!\!\!\!\!\! \!\!\!\!\!\! \!\!\!   \times \rho_{S_L,S_L}
+\Big(- \gamma_1 |\alpha_A|  +\gamma_2  \frac{\alpha_A^{*\, 2}}{ |\alpha_A|} +
\gamma_{12}  \nonumber  \\
  &&  \!\!\!\!\!\! \!\!\!\!\!\! \!\!\!\!\!\! \!\!\!    \times
\Big(\alpha_A|\alpha_A|  -\frac{\alpha_A}{|\alpha_A|} \Big) \Big)
\rho_{S_L,A_L}       +\Big(- \gamma_1 |\alpha_A|  \nonumber \\
  && \!\!\!\!\!\! \!\!\!\!\!\! \!\!\!\!\!\! \!\!\!    +\gamma_2
\frac{\alpha_A^{ 2}}{ |\alpha_A|} + \gamma_{12}   \Big(\alpha_A^*|\alpha_A|
-\frac{\alpha_A^*}{|\alpha_A|} \Big)   \Big)  \rho_{A_L,S_L}  \nonumber  \\
  &&  \!\!\!\!\!\! \!\!\!\!\!\! \!\!\!\!\!\! \!\!\!   +
(1+|\alpha_A|^2)\left(\gamma_0 -V_\mathrm{r}  \right) \rho_{A_L,A_L}
    \Big].
\end{eqnarray}

We remark that, in Eqs. (\ref{rate equation}) and as depicted in Fig.
\ref{FigTrans}, $\rho_{S_L,S_L}$, $\rho_{S_L,A_L}$, $\rho_{A_L,S_L}$, and
$\rho_{A_L,A_L}$ are, respectively, coupled only to themselves and $\rho_{E,E}$.


\begin{thebibliography}{99}
\bibitem{footnote}
We stress that in this case and for $\kappa_S=0$,
$V_{\mathrm{i}}$ is different from zero, leading to two different frequency components
in $\langle \sigma^x_k\rangle_t $ and $\langle \sigma^y_k\rangle_t $.
\end{thebibliography}
\end{document}